\documentclass[pra,aps,10pt,groupedaddress,twocolumn,floatfix]{revtex4-1}  
\usepackage{graphicx,color}
\usepackage{amsmath,amssymb,bm}
\usepackage{braket}		

\DeclareMathOperator{\Tr}{Tr}

\usepackage[plainpages=false,pdfpagelabels,colorlinks=true,linkcolor=red,urlcolor=blue,citecolor=blue,pdftitle={Titl}e,pdfauthor={},pdfdisplaydoctitle=true,pdfduplex=DuplexFlipLongEdge]{hyperref}

\definecolor{darkred}{rgb}{0.90,0.2,0.2}
\definecolor{darkgreen}{rgb}{0,0.60,.2}
\definecolor{darkblue}{rgb}{0.1,0.3,1}
\definecolor{grey}{cmyk}{0,0,0,0.25}
\definecolor{orange}{cmyk}{0,0.6,0.8,0}

\begin{document}
\title{Information measures for a local quantum phase transition:\\ Lattice fermions in a one-dimensional harmonic trap}

\author{Yicheng Zhang}
\author{Lev Vidmar}
\author{Marcos Rigol}
\affiliation{Department of Physics, The Pennsylvania State University, University Park, Pennsylvania 16802, USA}

\begin{abstract}
We use quantum information measures to study the local quantum phase transition that occurs for trapped spinless fermions in one-dimensional lattices. We focus on the case of a harmonic confinement. The transition occurs upon increasing the characteristic density and results in the formation of a band-insulating domain in the center of the trap. We show that the ground-state bipartite entanglement entropy can be used as an order parameter to characterize this local quantum phase transition. We also study excited eigenstates by calculating the average von Neumann and second Renyi eigenstate entanglement entropies, and compare the results with the thermodynamic entropy and the mutual information of thermal states at the same energy density. While at low temperatures we observe a linear increase of the thermodynamic entropy with temperature at all characteristic densities, the average eigenstate entanglement entropies exhibit a strikingly different behavior as functions of temperature below and above the transition. They are linear in temperature below the transition but exhibit activated behavior above it. Hence, at nonvanishing energy densities above the ground state, the average eigenstate entanglement entropies carry fingerprints of the local quantum phase transition.
\end{abstract}

\maketitle


\section{Introduction}

Since the early days of the exploration of strongly correlated many-body quantum systems with ultracold atoms in optical lattices~\cite{jaksch_98, greiner02, greiner_mandel_02b}, paradigmatic lattice models have been realized with the presence of additional inhomogeneous trapping potentials \cite{bloch08, cazalilla_citro_review_11}. Over the years~\cite{batrouni_rousseau_02, kashurnikov_prokofev_02, rigol_muramatsu_03, rigol_muramatsu_04may, kollath_schollwoeck_04, bergkvist_henelius_04, wessel_alet_04, rigol_muramatsu_04sept, rigol_muramatsu_05july, ma_yang_08, batrouni_krishnamurthy_08, rigol_batrouni_09, campostrini_vicari_09, pollet_prokofev_10, Spielman_10, campostrini_vicari_10b, campostrini_vicari_10c, mahmud_duchon_11, pollet_12, ceccarelli_torrero_13, angelone_campostrini_14, xu_rigol_15}, one of the central goals has been to understand how the results of measurements in inhomogeneous systems can be related to the phases and quantum phase transitions that occur in their homogeneous counterparts~\cite{sachdevbook}. In this work, we revisit this question in the context of recent progress in measuring the second Renyi entanglement entropy with ultracold atoms in optical lattices~\cite{islam_ma_15, kaufman_tai_16}. This has opened a new window for the exploration of many-body physics and quantum phase transitions~\cite{osterloh_amico_2002, osborne_nielsen_02, gu_deng_04} at the interface with quantum information theory~\cite{campostrini_vicari_10a, calabrese_mintchev_11, calabrese_mintchev_12a, vicari12, calabrese_ledoussal_15, alba_17, dubail_stephan_17, alba_calabrese_17, alba_calabrese_17b, dechiara_sanpera_17}.

Several properties of trapped systems are noticeably different from their homogeneous counterparts. For example, the quasimomentum distribution function of trapped noninteracting fermions at zero temperature does not exhibit the traditional step-function shape~\cite{rigol_muramatsu_04b}. For lattice bosons (fermions), Mott-insulating and superfluid (metallic) phases can coexist space separated in a trap~\cite{batrouni_rousseau_02, kashurnikov_prokofev_02, rigol_muramatsu_03, rigol_muramatsu_04may}. More importantly, the emergence of Mott-insulating domains in a trapped system does not follow the traditional quantum phase-transition paradigm, in which a global order parameter (the compressibility) vanishes in the insulating phase. In trapped systems, the global compressibility is nonzero no matter whether Mott domains are absent or present~\cite{batrouni_rousseau_02}. Instead, local order parameters, such as the fluctuations of the site occupation or local compressibilities \cite{batrouni_rousseau_02, rigol_muramatsu_03, wessel_alet_04}, need to be used to characterize the Mott domains. Hence, the term {\it local} quantum phase transition is more fitting to describe the formation of a Mott-insulating domain in a trapped system. Also, rather than phase diagrams, state diagrams are the proper way to describe trapped systems in the thermodynamic limit~\cite{rigol_muramatsu_03, rigol_batrouni_09, Spielman_10, angelone_campostrini_14}. In the state diagrams, the characteristic density $\rho$, which is the ratio between the particle number $N$ and the characteristic trap length $R$ [to be defined in Eq.~(\ref{def_H})], replaces the density used in the homogeneous case.

In this work, we study the behavior of quantum information measures when a band-insulating domain emerges in systems with noninteracting spinless fermions in the presence of a harmonic trap in one-dimensional lattices. We characterize both the ground state and excited states by using the von Neumann and second Renyi eigenstate entanglement entropies. In the ground state, a local quantum phase transition occurs at a critical characteristic density $\rho_c$. Below $\rho_c$, the site occupations are smaller than one in the entire lattice, while above $\rho_c$ a band-insulating domain with site occupation one forms in the center of the trap. We show that the entanglement entropy, upon a bipartition of the lattice in two equal parts, can be used as an order parameter for the local quantum phase transition. The parameter regime we focus on, namely, $\rho \sim \rho_c$, is complementary to that accessed recently in a conformal field theory study~\cite{dubail_stephan_17}.

For excited eigenstates, we study the average von Neumann and second Renyi entropies and compare them with the mutual information and thermodynamic entropies of thermal states. It was recently proved that, for translationally-invariant fermionic quadratic models, the half-system bipartite entanglement entropy of typical eigenstates is smaller than the thermodynamic entropy of thermal states at the same mean energy (the difference is extensive in the system size)~\cite{vidmar_hackl_17}. Here we not only show that the same holds for harmonically trapped quadratic systems (i.e., in systems that are not translationally invariant), but also that the average eigenstate entanglement entropy (with Gibbs weights) may exhibit a completely different functional dependence on the characteristic density and the temperature than the thermodynamic entropy. We note that the nature of the inhomogeneity considered here is different from that in previous studies in which translational invariance was broken by means of diagonal disorder~\cite{huang_moore_14} and random long-range hoppings~\cite{liu_chen_17}.

Our work reveals a particularly interesting behavior of the average eigenstate entanglement entropy as a function of temperature. At low temperatures, it increases linearly with temperature below $\rho_c$ (similarly to the thermodynamic entropy of the translationally invariant metallic counterpart), but exhibits an activated behavior above $\rho_c$. Hence, at nonvanishing energy densities above the ground state, the average eigenstate entanglement entropy allows one to identify the presence of the ground-state local quantum phase transition. This is in stark contrast with the thermodynamic entropy, which is linear in temperature independently of the value of $\rho$. The latter is the result of the many-body energy spectrum being gapless irrespective of the value of $\rho$.

The presentation is organized as follows: In Sec.~\ref{sec2}, we introduce the model and the entanglement measures used to characterize it. In Sec.~\ref{sec3}, we study the properties of the ground state, while Sec.~\ref{sec4} is devoted to the study of properties at nonvanishing energy densities above the ground state. A summary of the results is presented in Sec.~\ref{sec5}. We discuss details of the numerical calculations in Appendixes~\ref{app0}--\ref{app4}.

\section{Model and information measures} \label{sec2}

We consider spinless fermions in a one-dimensional lattice with $L$ (even) sites, described by the Hamiltonian
\begin{equation} \label{def_H}
\hat H = -t \, \sum_{l=0}^{L-1} (\hat f_l^\dagger \hat f_{l+1} + {\rm H.c.}) \, + \frac{t}{R^2}\sum_{l=0}^{L-1} \left( l - \frac{L-1}{2} \right)^2 \hat n_l \, ,
\end{equation}
where $\hat f_l$ is the fermion annihilation operator at site $l$, and $\hat n_l = \hat f_l^\dagger \hat f_l$. We set the unit of energy $t=1$ and the lattice spacing to one. The strength of the harmonic confining potential is determined by the parameter $R$, which is a characteristic length for the trapped lattice system. We therefore study the properties of the system as a function of the characteristic density $\rho = N/R$~\cite{rigol_muramatsu_04b}. To have a vanishing density at the edges of the lattice, in the ground state calculations we take $L\gtrsim 4R$ [see Fig.~\ref{fig1}(b)], while in the excited state calculations we take $L\gtrsim 8R$.

Even though the model is quadratic, an analytic solution for the single-particle energy eigenstates is only available at low and high energies~\cite{rey_pupillo_05}. At intermediate energies, relevant to the formation of the band-insulating domain in the center of the trap, the properties of the single-particle eigenstates can be studied numerically. Using local observables, such as the site occupations and their fluctuations, the critical characteristic density for the formation of the band-insulating domain in the center of the trap was shown to be $\rho_c \approx 2.6$~\cite{rigol_muramatsu_04b}. At the corresponding Fermi energy, a semiclassical Wenzel-Kramers-Brillouin (WKB) approximation shows that the single-particle density of states exhibits a logarithmic singularity~\cite{hooley_quintanilla_04}. These findings will be further discussed in this work.

For single-particle eigenenergies greater than the Fermi energy corresponding to $\rho_c$, the eigenstates become doubly degenerate (they are even or odd upon lattice reflection). We weakly break this degeneracy by modifying $R^{-2} \to R^{-2} (1\mp \eta)$ in Eq.~(\ref{def_H}) for sites $l \lessgtr (L-1)/2$, with $\eta \ll 1$. We choose $\eta = 1/N_c = 1/(R \rho_c)$, with the value of $\rho_c$ to be determined later [see Eq.~\ref{def_rhoc}].

We study the entanglement entropy of the many-body eigenstates of $\hat H$ in Eq.~(\ref{def_H}) for bipartitions of the system into equal subsystems $A$ and $\bar A$. For a many-body eigenstate $|m\rangle$, the reduced density matrix of the subsystem $A$ is $\hat \rho(m) = {\rm Tr}_{\bar A} (|m\rangle \langle m|)$. We are interested in the second Renyi entropy, $S_{\rm n}(m)=\frac{1}{1-{\rm n}}\ln{\Tr [ \hat\rho(m)^{\rm n} ] }$ for ${\rm n}=2$, and in the von Neumann entropy, $S_{\rm vN}(m)=-\Tr [ \hat \rho(m)\ln\hat\rho(m) ]$, which is the limit ${\rm n}\to1$ of $S_{\rm n}$. For many-body eigenstates of noninteracting fermions, the reduced density matrix $\hat \rho(m)$ has a Gaussian form. $\hat \rho(m)$ can be obtained from the one-body (covariance) correlation matrix $F_{ij}(m)=\langle m | \hat f_i^\dagger \hat f_j | m\rangle$ defined on the sites $i,j$ of subsystem $A$~\cite{peschel03}. The entanglement entropies are then unique functions of the eigenvalues $\lambda_j$ of $F(m)$, which we diagonalize numerically. Specifically, the von Neumann entanglement entropy is
\begin{equation}\label{def_S}
 S_{\rm vN}(m) = - \sum_{j=1}^{L/2} \Big[ \lambda_j \ln \lambda_j + (1-\lambda_j) \ln (1-\lambda_j) \Big] \, ,
\end{equation}
and the second Renyi entanglement entropy is
\begin{equation}\label{def_S_n}
 S_2(m) =-\sum_{j=1}^{L/2}\ln\left[(1-\lambda_j)^2+\lambda_j^2\right] \, .
\end{equation}

In the continuum, the limit $\rho\to0$ in the lattice, the behavior of the Renyi entanglement entropies of the ground state of trapped noninteracting fermions is well understood. The leading term scales as $\ln N$~\cite{calabrese_mintchev_11, calabrese_mintchev_12a}. It has been shown that the prefactor in the leading term is $(1+{\rm n}^{-1})/12$, where ${\rm n}$ is the order of the Renyi entanglement entropy,  i.e., it is identical to the one in homogeneous systems with open boundaries~\cite{vicari12}. In lattice systems with nonvanishing values of $\rho$, the systems of interest here, such a scaling is in general no longer valid.

We also study entanglement properties of excited eigenstates. We calculate the average eigenstate entanglement entropy $\bar S_{\rm n}(T)$ at a given mean energy density as
\begin{equation}\label{def_S_T}
\bar S_{\rm n}(T) = \frac{\sum_m{S_{\rm n}(m)e^{-E_m/T}}}{\sum_m{e^{-E_m/T}}}\, ,
\end{equation}
where ${\rm n} = {\rm vN}$ or ${\rm n} = 2$, and $E_m$ is the energy of eigenstate $|m\rangle$. The temperature $T$ sets the mean energy density of the eigenstates involved in the average, and the summation runs over eigenstates with a fixed $N$, i.e., it corresponds to a canonical ensemble average. In the numerical calculation of Eq.~(\ref{def_S_T}), we discard eigenstates with a relative weight $\exp[-(E_m-E_{\rm GS})/T]$, where $E_{\rm GS}$ is the ground-state energy, below a threshold value $\exp[{-\Lambda}]$. We use $\Lambda = 30$, which yields a negligibly small numerical error, as shown in Fig.~\ref{figapp1} of Appendix~\ref{app1}. In Eq.~(\ref{def_S_T_GE}) of Appendix~\ref{app2}, we also extend Eq.~(\ref{def_S_T}) to compute grand canonical ensemble averages. We find that the results for the canonical and grand canonical ensemble averages are very similar, see Fig.~\ref{figapp2} in Appendix~\ref{app2}. Therefore, in what follows we only discuss the results for the canonical ensemble average.

We compare the average eigenstate entanglement entropy in Eq.~(\ref{def_S_T}) to properties of (mixed) thermal states $\hat \rho(T) = e^{-(\hat H-\mu \hat N)/T}/{\rm Tr}\{ e^{-(\hat H - \mu \hat N)/T} \}$ at the same temperature and average particle number $N$ (used to determine $\mu$). In particular, we study the quantum mutual information
\begin{equation}\label{def_I}
I(T) = 2 S_{\rm vN}(T) - S_{\rm GE}(T)\, ,
\end{equation}
where $S_{\rm vN}(T)$ is the von Neumann entropy of the reduced density matrix of $\hat \rho(T)$, and $S_{\rm GE}(T)$ is the thermodynamic (grand canonical ensemble) entropy. The calculation of $S_{\rm vN}(T)$ can be done using Eq.~(\ref{def_S}), with $\lambda_j$ being the eigenvalues of the one-body correlations matrix $F(T)$ with matrix elements $F_{ij}(T) = {\rm Tr} \{ \hat \rho(T) \hat f_i^\dagger \hat f_j \}$. The thermodynamic entropy $S_{\rm GE}(T) =-\Tr[\hat\rho(T)\ln{\hat\rho(T)}]$ can be obtained by using the expression
\begin{equation}
S_{\rm GE}(T) = - \sum_{j=1}^L \Big[ n_j \ln n_j + (1-n_j) \ln (1-n_j) \Big] \, ,
\end{equation}
where $n_j =\left[1+e^{(\varepsilon_j-\mu)/T}\right]^{-1}$, and $\varepsilon_j$ is the single-particle eigenenergy. Note that the quantum mutual information is not an entanglement measure, but it quantifies the amount of correlations between the two subsystems. For thermal states, it was proved that $I(T)$ follows an area law with the system size~\cite{groisman_popescu_05, wolf_verstraete_08}.

\section{Local quantum phase transition} \label{sec3}

\begin{figure}[!t]
\begin{center}
\includegraphics[width=0.99\columnwidth]{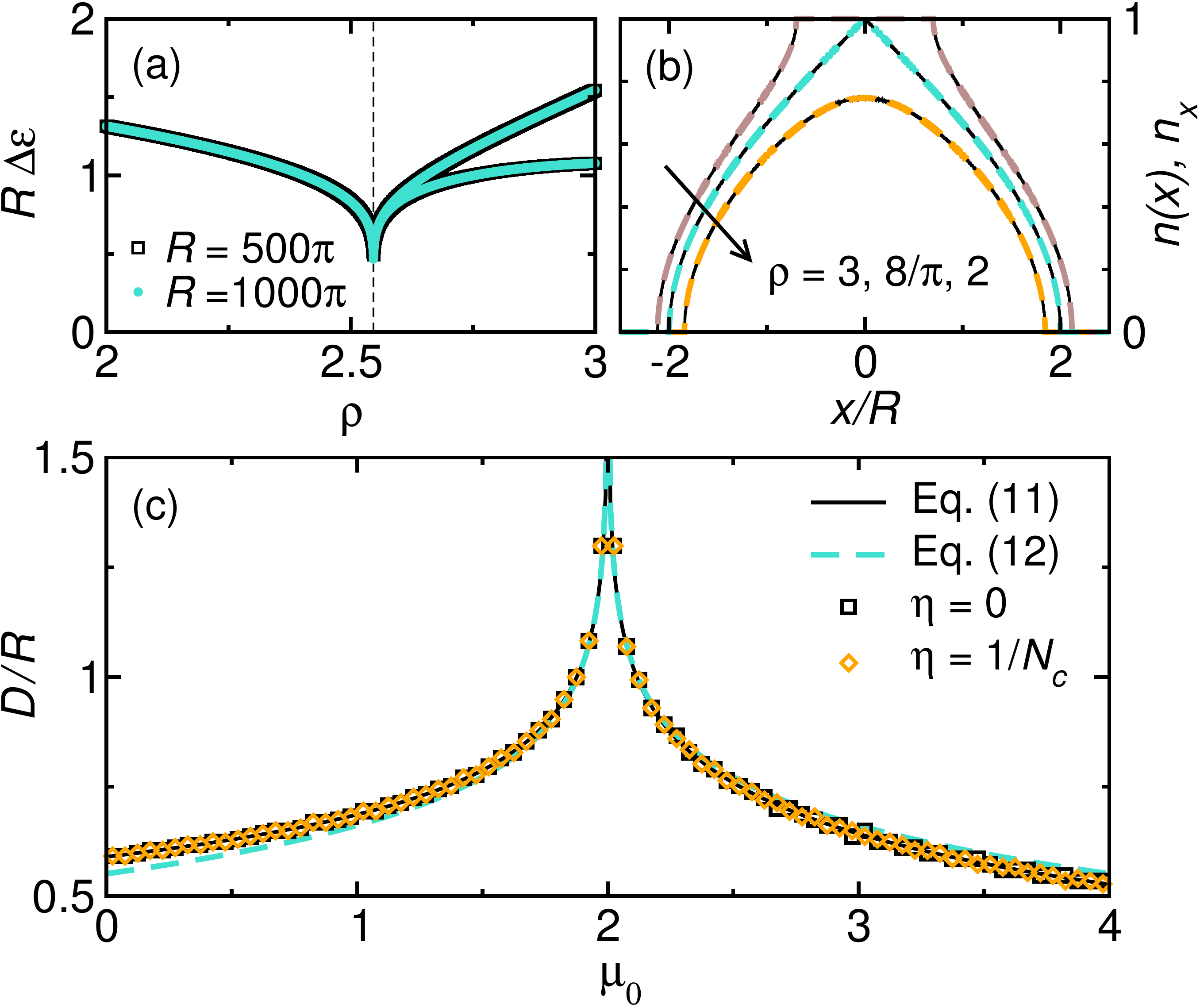}
\caption{Properties of one-dimensional lattice fermions in a harmonic trap. (a) Second derivative of the total energy vs $\rho$, $\bar E''(\rho) = R \Delta \varepsilon_N$, which is proportional to the level spacing $ \Delta \varepsilon_N = \varepsilon_{N+1} - \varepsilon_N$ of the single-particle energy spectrum of $\hat H$ in Eq.~(\ref{def_H}). The dashed line denotes the critical characteristic density $\rho_c=8/\pi$ [see Eq.~\eqref{def_rhoc}]. We choose the parity-breaking parameter $\eta = 1/N_c$ throughout the paper [see text after Eq.~\eqref{def_H}]. (b) Site occupations at (from top to bottom) $\rho=3$, $\rho=\rho_c=8/\pi$, and $\rho=2$. Solid lines are exact numerical results $n_x$ (with integer $x$) and the overlapping dashed lines are the LDA results $n(x)$ from Eq.~(\ref{def_n_x}). (c) Single-particle density of states $D(\mu_0)/R$, where $\mu_0$ is the single-particle energy (the Fermi energy of the many-body ground state). The solid and dashed lines are the LDA results from Eq.~(\ref{DOS}) and the corresponding leading term from Eq.~(\ref{def_logD}), respectively. The symbols are the numerical exact diagonalization results with $(\eta = 1/N_c)$ and without $(\eta = 0)$ parity breaking. The results in panels (b) and (c) were obtained for $R=500\pi$.} 
\label{fig1}
\end{center}
\end{figure}

We first focus on the properties of the ground state of $\hat H$ in Eq.~(\ref{def_H}). We define the total energy density as $\bar E = E/R = \sum_{j=1}^N \varepsilon_j/R$. The discrete second derivative of $\bar E$ with respect to $\rho$ is $\bar E'' = R \Delta \varepsilon_N$, where $\Delta \varepsilon_N = \varepsilon_{N+1} - \varepsilon_N$ is the level spacing of the single-particle spectrum of $\hat H$. Results for $\bar E''(\rho)$ vs $\rho$ are shown in Fig.~\ref{fig1}(a) for two values of $R$ (i.e., for two system sizes). They exhibit a perfect collapse. These results provide two important insights. First, the scaling $\Delta \varepsilon_N \approx \mbox{const.}/R$ indicates that the system is gapless in the thermodynamic limit ($R\rightarrow\infty$) for all values of $\rho$ under investigation. Second, $\bar E''$ vs $\rho$ at $\rho \approx 2.6$ appears to be nonanalytic, a behavior that is typical of a quantum phase transition~\cite{sachdevbook}. We show in what follows that $\bar E''(\rho)$ is indeed nonanalytic at $\rho=\rho_c$, a point at which a band-insulating domain forms in the center of the system [see Fig.~\ref{fig1}(b)]. The transition in this case can be regarded as a local quantum phase transition, as argued earlier.

\subsection{Local density approximation}

Here, we use the local density approximation (LDA) to describe two features: (i) the formation of the band-insulating domain in the center of the trap~\cite{rigol_muramatsu_04b, angelone_campostrini_14}, and (ii) the divergence of the single-particle density of states~\cite{hooley_quintanilla_04}.

Within the LDA, one constructs an effective local chemical potential at the (continuous) position $x$, which is the distance from the center of the trap,
\begin{equation}
\mu(x)=\mu_0-\frac{x^2}{R^2} \, ,
\end{equation} 
where $\mu_0$ is the chemical potential (Fermi energy), with $\mu_0 > -2$ to have nonzero site occupations. Then, for each value of $x$, the system is treated as a homogeneous one with chemical potential $\mu(x)$, e.g., $n(x)=k_F(x)/\pi$ with $k_F(x)$ and $\mu(x)$ related through $\mu(x)=-2\cos [k_F(x)]$.

The site occupations $n(x)$ vanish for all $|x| > x_0$, where $x_0 = R\sqrt{\mu_0+2}$. For $\mu_0 < 2$, $n(x) < 1$ for all $x$ since
\begin{equation} \label{def_n_x_0}
 n(x) = \frac{1}{\pi}\arccos\left[-\frac{\mu(x)}{2}\right] \, .
\end{equation}
On the other hand, for $\mu_0 \geq 2$, a band-insulating domain is present for $|x| < x_1$, where $x_1 = R\sqrt{\mu_0-2}$. The site-occupation distribution in the trap is then
\begin{equation}
n(x)= \Bigg \{ \begin{tabular} {ccl} $1,$ && $|x|<x_1$\\ 
$\frac{1}{\pi}\arccos\left(-\frac{\mu(x)}{2}\right),$&&$x_1 < |x| < x_0.$ \end{tabular}
\label{def_n_x}
\end{equation}

Next, we relate the total number of particles in the trap, $N=\int_{-x_0}^{x_0} n(x)dx$, to the chemical potential $\mu_0$. The integration results in $N = R \, g(\mu_0)$~\cite{batrouni_krishnamurthy_08, xu_rigol_15}, where $g(\mu_0)$ is given by Eq.~(\ref{def_rho_mu0}) in Appendix~\ref{app0}, for $\mu_0\leq2$. This explains why the characteristic density $\rho = N/R$ needs to be kept constant when taking the thermodynamic limit in a trapped system~\cite{rigol_muramatsu_03, batrouni_krishnamurthy_08, rigol_batrouni_09, angelone_campostrini_14, xu_rigol_15}. Figure~\ref{fig1}(b) shows the site-occupation distribution $n(x)$ predicted by the LDA at three characteristic densities $\rho$ as a function of $x/R$. They are indistinguishable from the exact numerical results $n_x=\langle \hat{n}_x\rangle$. Figure~\ref{figapp0} in Appendix~\ref{app0} shows site-occupation profiles for small values of $\rho$ compared against the predictions of the LDA in the continuum limit.

The band-insulating domain emerges in the center of the trap when the occupation there becomes one, i.e., when $\mu_0=2$. This allows us to obtain the critical particle number $N_c=\int_{-x_0}^{x_0} n_c(x)dx = 8R/\pi$, so that the critical characteristic density is
\begin{equation} \label{def_rhoc}
\rho_c=\frac{N_c}{R}=\frac{8}{\pi}\, .
\end{equation}
This value matches the point at which the second derivative of the total energy density appears to behave nonanalytically in Fig.~\ref{fig1}(a).

The LDA also allows one to calculate the single-particle density of states, $D(\mu_0) = dN/d\mu_0$. It gives
\begin{equation}
D(\mu_0) =\frac{2R}{\pi \sqrt{2-\mu_0}}\bigg \{
\begin{tabular} {cc} $K(t), $ & $-2\leq\mu_0<2$\\$K(t)-F(t^{-1/2}|t),$ & $\mu_0>2,$ \end{tabular}
\label{DOS}
\end{equation}
where $t=(\mu_0+2)/(\mu_0-2)$, and $F(t\big|m)$ [$K(t)$] is the elliptic integral (complete elliptic integral) of the first kind~\cite{arfken_weber_08}. A divergence of the density of states occurs at $\mu_0 = 2$, which corresponds to $\rho_c$ in Eq.~(\ref{def_rhoc}). Beyond this point ($\varepsilon_j > 2$), single-particle states become localized at either the left or the right side of the center of the trap, due to the Bragg reflection~\cite{rigol_muramatsu_04b, hooley_quintanilla_04}. The exact numerical results for the density of states shown in Fig.~\ref{fig1}(c) are in perfect agreement with those obtained evaluating Eq.~(\ref{DOS}). They are also identical when $\eta=1/N_c$ and $\eta=0$, i.e., the anisotropy used in our calculations almost does not change the density of states. Close to $\mu_0 = 2$, one can expand Eq.~(\ref{DOS}) to obtain the leading-order term
\begin{equation} \label{def_logD}
\lim_{\mu_0\to 2}D(\mu_0) = \frac{R}{2\pi} \left(6\ln 2-\ln\left|\mu_0-2\right| \right) \, ,
\end{equation}
which confirms that the divergence is logarithmic, as advanced in Ref.~\cite{hooley_quintanilla_04}.

\subsection{Order parameter: Entanglement entropies}  \label{sgs}

Our analysis so far has revealed that, for $\rho > \rho_c$ in the thermodynamic limit, there are lattice sites in the center of the trap with site occupation one. The correlations $\langle \hat f_i^\dagger \hat f_j \rangle$ across those sites are therefore zero (in the absence of degeneracies in the single-particle spectrum, which is our case). Those sites with occupation one split the $L$-by-$L$ one-body correlation matrix $F$ into two disconnected blocks. The eigenvalues within each block (relevant to obtain $S_{\rm vN}$ and $S_{\rm n}$) are those of a pure state. The entanglement entropies $S_{\rm vN}$ and $S_{\rm n}$ therefore must vanish for $\rho > \rho_c$, which makes them suitable candidates for the order parameter of the local quantum phase transition.

\begin{figure}[!t]
\begin{center}
\includegraphics[width=0.99\columnwidth]{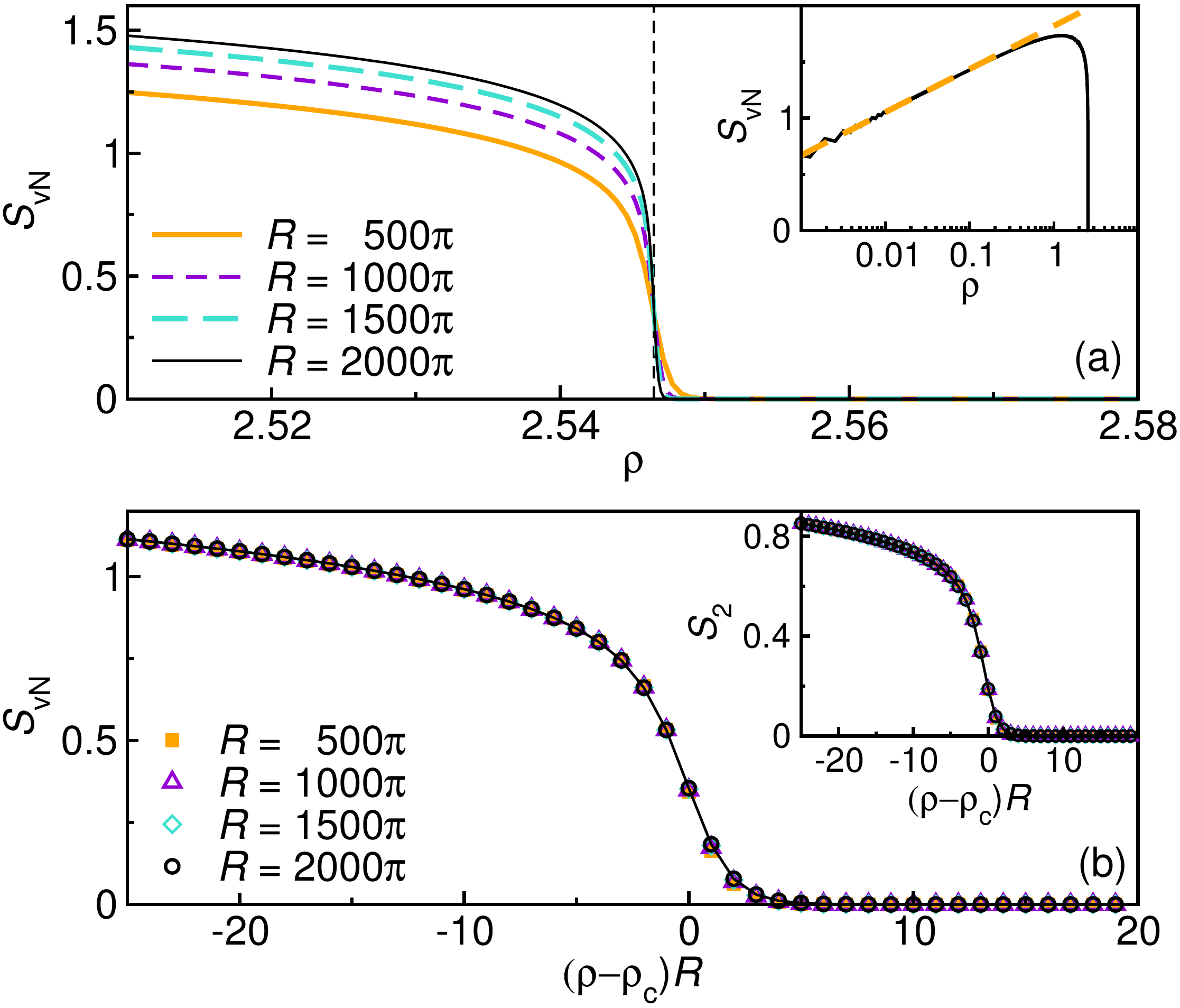}
\caption{Ground-state entanglement entropy. (a) von Neumann entanglement entropy $S_{\rm vN}$ vs $\rho$ about the critical point (main panel), and in a wider range (inset, results for $R=500\pi$). The vertical dashed line in the main panel denotes the critical point $\rho_c=8/\pi$. The dashed line in the inset is $S_{\rm vN} = (1/6) \left[\ln{N} + \ln{8} + 1.485\right]$ from Ref.~\cite{vicari12}. (b) Rescaling of the $x$ axis, $\rho \to (\rho - \rho_c)R$, which results in data collapse about the critical point. We plot $S_{\rm vN}$ (main panel) and $S_2$ (inset) for different values of $R$, as indicated in the legend.} 
\label{fig2}
\end{center}
\end{figure}

The main panel of Fig.~\ref{fig2}(a) shows $S_{\rm vN}$ as a function of $\rho$ across the local quantum phase transition, for four values of $R$. In finite systems, $S_{\rm vN}$ vanishes when $\rho > \rho_c$. Moreover, the curves for different values of $R$ (for sufficiently large $R$) cross at $\rho_c$ [see the vertical line in Fig.~\ref{fig2}(a)], and they become sharper with increasing $R$. This observation is consistent with a vanishing $S_{\rm vN}$ for any $\rho > \rho_c$ in the thermodynamic limit. We note that $S_{\rm vN}$ is not an extensive quantity~\cite{amico_fazio_08,eisert_cramer_10}, so it is fitting that this is the kind of order parameter that one needs to characterize the local quantum phase transition undergone by these systems. 

The inset in Fig.~\ref{fig2}(a) shows that $S_{\rm vN}$ is nonzero for all nonzero $\rho < \rho_c$ (the same holds true for $S_2$, not shown). The dashed line depicts the analytical result in the continuum limit~\cite{vicari12}, $S_{\rm vN} = (1/6) \left[\ln{N} + \ln{8} + y_1\right]$, with $y_1 \approx 1.485$. It provides an excellent description for the results in the lattice for small values of the characteristic density, $\rho \lesssim 0.5$.

Finally, we use the value of $\rho_c$ predicted by the LDA, Eq.~(\ref{def_rhoc}), to study the scaling of the entanglement entropies across the transition. The main panel (inset) of Fig.~\ref{fig2}(b) shows a perfect data collapse for $S_{\rm vN}$ ($S_2$) vs $(\rho - \rho_c)R$ for four values of $R$. This suggest that
\begin{equation}
S_{\rm n} = \mathcal{F}_{\rm n} \left[ (\rho-\rho_c) R\right] = \mathcal{F}_{\rm n} \left[ N-N_c\right] \,
\end{equation}
is a universal scaling function that describes the corresponding entanglement entropy across the local quantum phase transition.

From our results in this section we conclude that, through a scaling analysis, experimental measurements of entanglement entropies in finite systems can enable the location of the local quantum phase transition in the thermodynamic limit. The result obtained from such an analysis will be much more accurate than those obtained by using local observables such as the site occupations and their fluctuations. 

\section{Finite temperature properties} \label{sec4}

\subsection{Thermal states: Mutual information and thermodynamic entropy}
\begin{figure}[!t]
\begin{center}
\includegraphics[width=0.99\columnwidth]{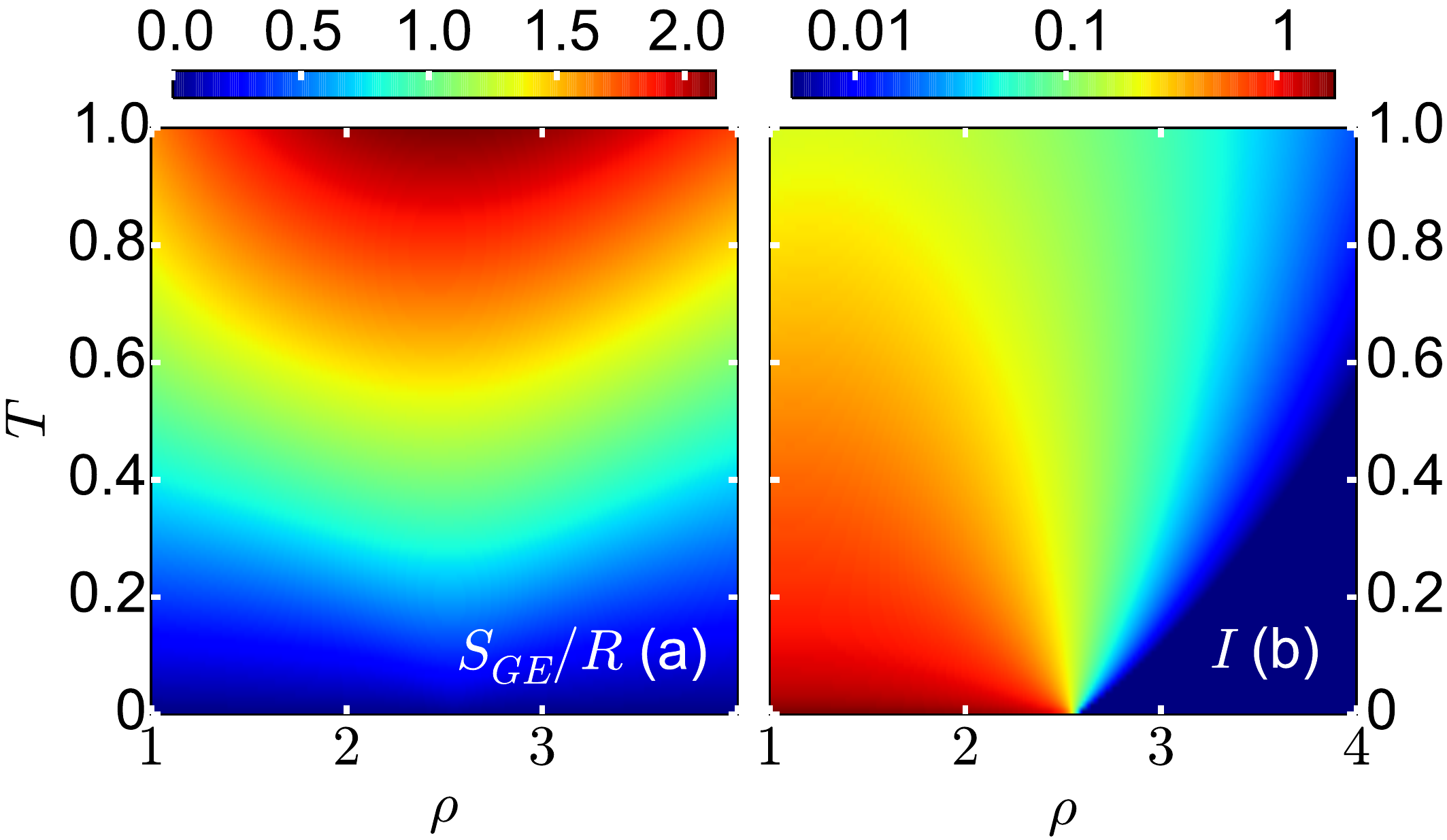}
\caption{Numerical results for: (a) the thermodynamic entropy density, $S_{\rm GE}/R$, and (b) the mutual information, $I$, as a function of $\rho$ and $T$ for trapped systems with $R=125\pi$.} 
\label{fig3}
\end{center}
\end{figure}

Here we study the properties of the trapped system in thermal equilibrium. The density plots in Figs.~\ref{fig3}(a) and~\ref{fig3}(b) show the thermodynamic entropy density $S_{\rm GE}(T,\rho)/R$ and the mutual information $I(T,\rho)$, respectively, as a function of $\rho$ and the temperature. 

\begin{figure}[!b]
\begin{center}
\includegraphics[width=0.99\columnwidth]{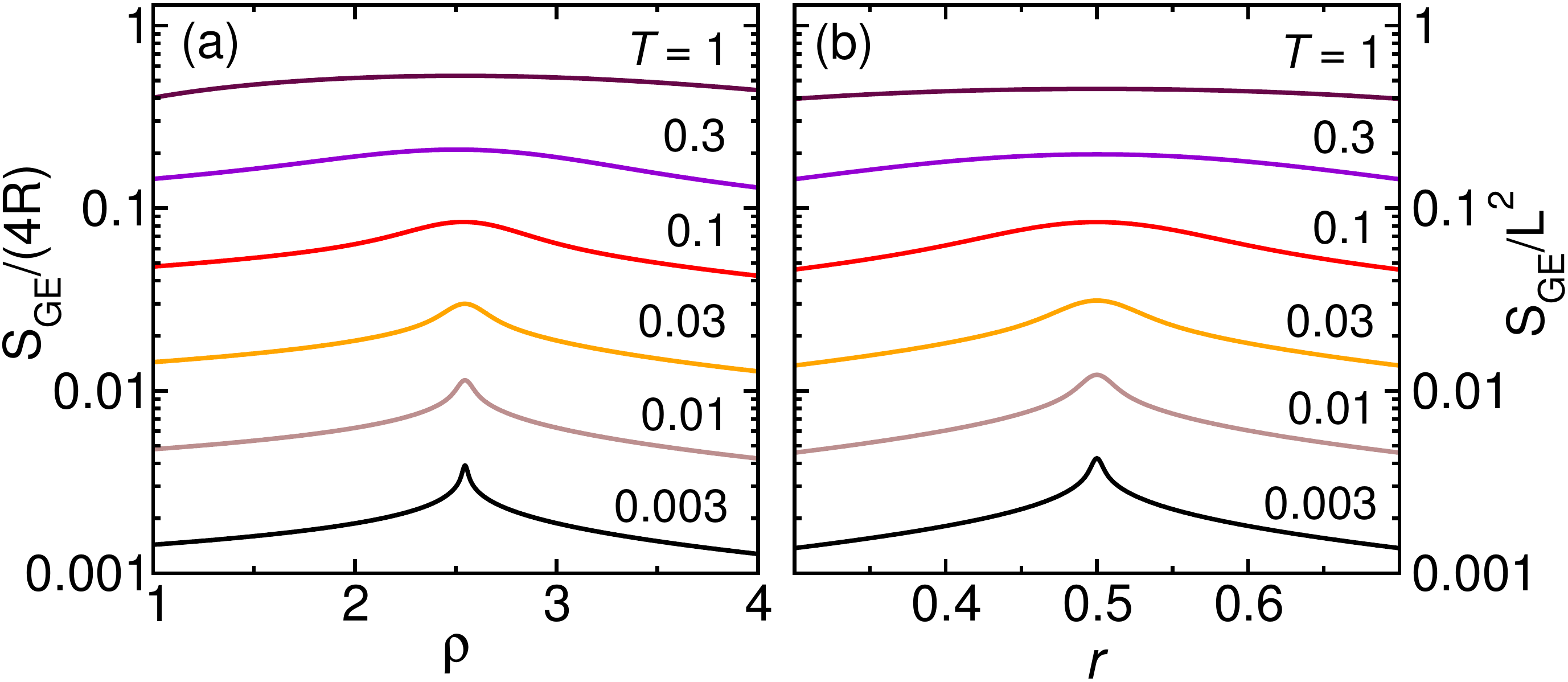}
\caption{Thermodynamic entropy density $S_{\rm GE}/{\cal N}$. (a) Entropy density of harmonically trapped fermions in a one-dimensional lattice, Eq.~(\ref{def_H}), with ${\cal N} = 4R$, as a function of the characteristic density $\rho = \langle N\rangle/R$ for $R=250\pi$. (b) Entropy density of fermions in a translationally invariant square lattice, Eq.~(\ref{def_H_2D}), with ${\cal N} = L^2$, as a function of the site occupation $r=\langle N\rangle/L$ for $L=2000$.} 
\label{fig4}
\end{center}
\end{figure}

The most prominent characteristic of $S_{\rm GE}$ is that it is maximal about $\rho_c$. This is better seen in Fig.~\ref{fig4}(a), where we plot $S_{\rm GE}$ vs $\rho$ for six temperatures (notice the log scale on the y axis). At low temperatures, a sharp peak can be seen about $\rho_c$. With increasing $T$, $S_{\rm GE}$ increases for all values of $\rho$, and the peak broadens and eventually disappears. The presence of a sharp peak at low temperatures is a direct consequence of the divergence of the density of states at the Fermi energy corresponding to $\rho_c$ [see Fig.~\ref{fig1}(c) and Eq.~(\ref{def_logD})], and could also be used to locate the local quantum phase transition in our trapped one-dimensional systems at low temperatures.

We should, however, mention that a similar peak can be seen in the entropy of noninteracting fermions in the translationally invariant square lattice with Hamiltonian
\begin{equation} \label{def_H_2D}
H_{\rm 2D}=-t \, \sum_{\langle{\bf i},{\bf j}\rangle} \left( \hat f_{{\bf i}}^\dagger \hat f_{\bf j} + {\rm H.c.} \right) \, ,
\end{equation}
where $\langle{\bf i},{\bf j}\rangle$ stands for nearest-neighbor sites. Such a system also exhibits a logarithmic divergence of the density of states at half filling ($\mu_0=0$). The entropy density for noninteracting fermions in the square lattice (with $L^2$ sites) is depicted in Fig.~\ref{fig4}(b). It can be seen to be qualitatively, and quantitatively, similar to that of the one-dimensional trapped system [whose number of occupied sites is about $4R$; see Fig.~\ref{fig1}(b)]. Hence, while a sharp peak in the entropy density does not necessarily indicate a local quantum phase transition (there is none in the two-dimensional system), it does for our trapped one-dimensional system.

\begin{figure}[!t]
\begin{center}
\includegraphics[width=0.99\columnwidth]{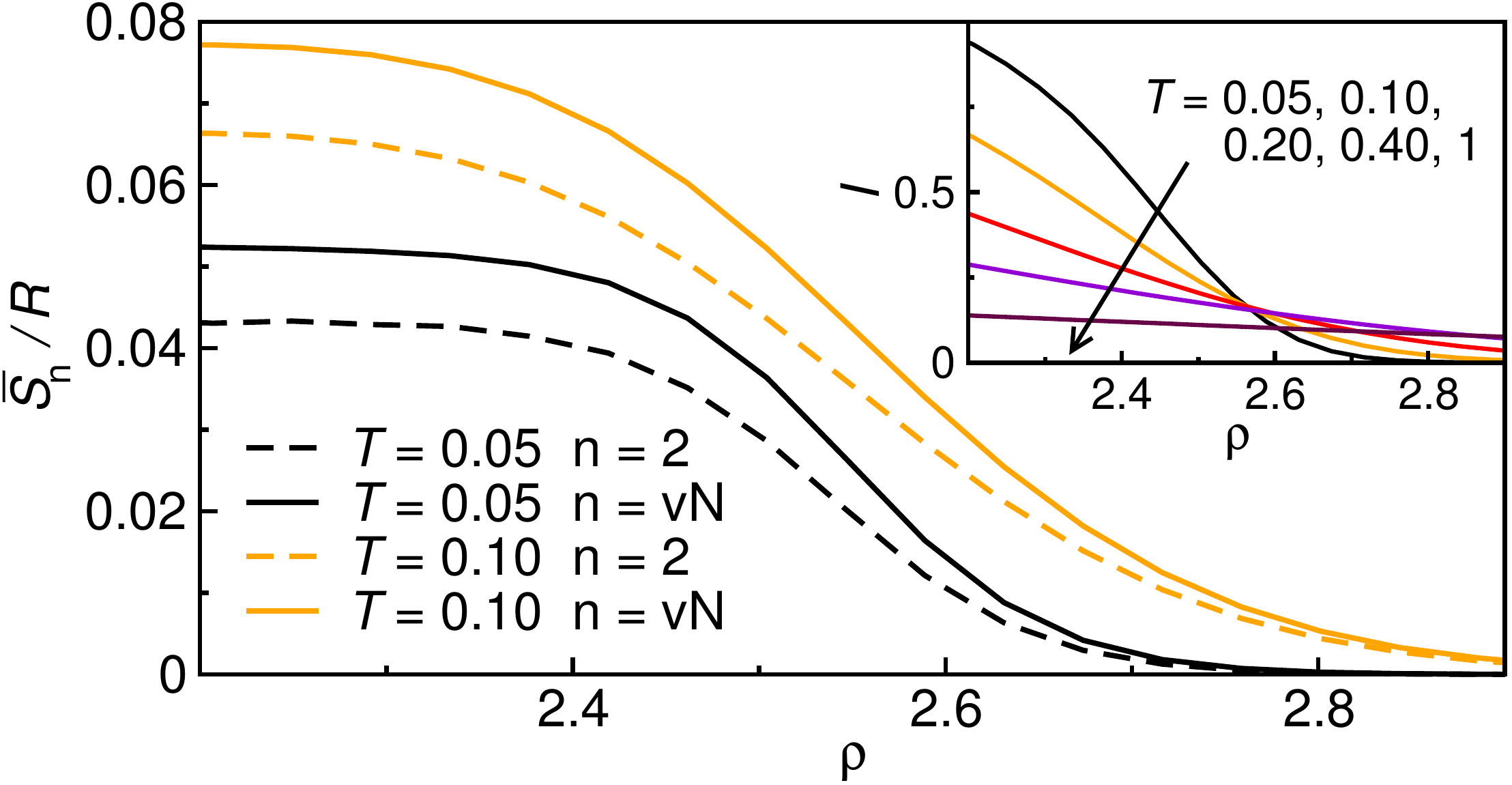}
\caption{(main panel) Average von Neumann entanglement entropy $\bar S_{\rm vN}/R$ (solid lines) and  second Renyi entanglement entropy $\bar S_{2}/R$ (dashed lines) in eigenstates of the Hamiltonian plotted as functions of $\rho$ for two temperatures. The inset shows the mutual information $I$ vs $\rho$ for five temperatures. All results were obtained for $R=7.5\pi$.} 
\label{fig5}
\end{center}
\end{figure}

On the other hand, at low temperatures, the behavior of the mutual information $I$ [Fig.~\ref{fig3}(b)] is qualitatively similar to that of the ground-state entanglement entropy. Namely, it is nonvanishing for $\rho < \rho_c$, and it is vanishingly small for $\rho > \rho_c$. Upon increasing $T$, the fingerprints of the local quantum phase transition disappear because $I(\rho < \rho_c)$ decreases (correlations between the two halves of the system decrease, an expected effect of the temperature) and $I(\rho>\rho_c)$ slightly increases (correlations between the two halves slightly increase because of the ``melting'' of the band-insulating domain), leading to a featureless structure at high temperatures. This is better seen in the inset of Fig.~\ref{fig5}, in which we plot the mutual information vs $\rho$ for five temperatures.

We note that, unlike $S_{\rm GE}$, the mutual information exhibits an area-law scaling (it is not divided by $R$). This is the result of the prefactor in the extensive part of $S_{\rm vN}(T)$ being identical to that of $S_{\rm GE}(T)/2$, see Eq.~(\ref{def_I}). One of the main goals of the next section is to show that this observation does not hold for the entanglement entropy of excited eigenstates at the same energy density. In particular, the prefactor in the extensive part of $S_{\rm vN}(m)$ for the overwhelming majority of the many-body eigenstates $|m\rangle$ at a given energy density can be considerably smaller than that in $S_{\rm vN}(T)$ at the same energy density.

\subsection{Average eigenstate entanglement entropy} \label{sec4b}

We now turn our focus to the average entanglement entropies of excited many-body eigenstates of the Hamiltonian~\eqref{def_H}, at mean energy densities that correspond to nonzero temperatures $T$. We perform large-scale numerical calculations to compute the weighted averages $\bar S_{\rm vN}$ and $\bar S_{2}$, defined in Eq.~(\ref{def_S_T}). Note that, since there are exponentially large numbers of eigenstates involved in the calculations, the values of $R$ accessible to us in this section are much smaller than in the previous sections. Still, they allow us to extract low-temperature properties in the thermodynamic limit.

The main panel of Fig.~\ref{fig5}(a) shows results for $\bar S_{\rm vN}$ and $\bar S_{2}$ as functions of $\rho$ for two temperatures. In contrast to the mutual information $I$, plotted in the inset of Fig.~\ref{fig5} and in Fig.~\ref{fig3}, $\bar S_{\rm vN}$ and $\bar S_{2}$ increase with increasing temperature for all values of $\rho$. This is understandable because, at nonzero mean energy densities above the ground state, the overwhelming majority of the many-body eigenstates are expected to exhibit a volume-law scaling with the system size, with a prefactor that increases with temperature. We explicitly verify the volume law scaling in Appendix~\ref{app3} for $\rho=2$ and $3$. Hence, at a fixed temperature $T$ and characteristic density $\rho$, we fit the numerical results with the ansatz
\begin{equation} \label{def_S_T_fit}
 \bar S_{\rm n}(T,\rho) = s_{\rm n}(T,\rho) \, R + \delta_{\rm n}(T,\rho),
\end{equation}
and extract the prefactor in the extensive part $s_{\rm n}$, i.e., the average eigenstate entanglement entropy density. Since $\rho$ needs to be kept fixed when taking the thermodynamic limit, the leading term in Eq.~(\ref{def_S_T_fit}) can also be written as $\left[ s_{\rm n}(T,\rho)/\rho \right] N$, making explicit the linear dependence on the number of particles. The numerical results in Fig.~\ref{figapp3} of Appendix~\ref{app3} agree perfectly with the functional form given in Eq.~(\ref{def_S_T_fit}). As expected, the von Neumann entropy density $s_{\rm vN}$ is always above the second Renyi entropy density $s_2$.

\begin{figure*}[!t]
\begin{center}
\includegraphics[width=1.99\columnwidth]{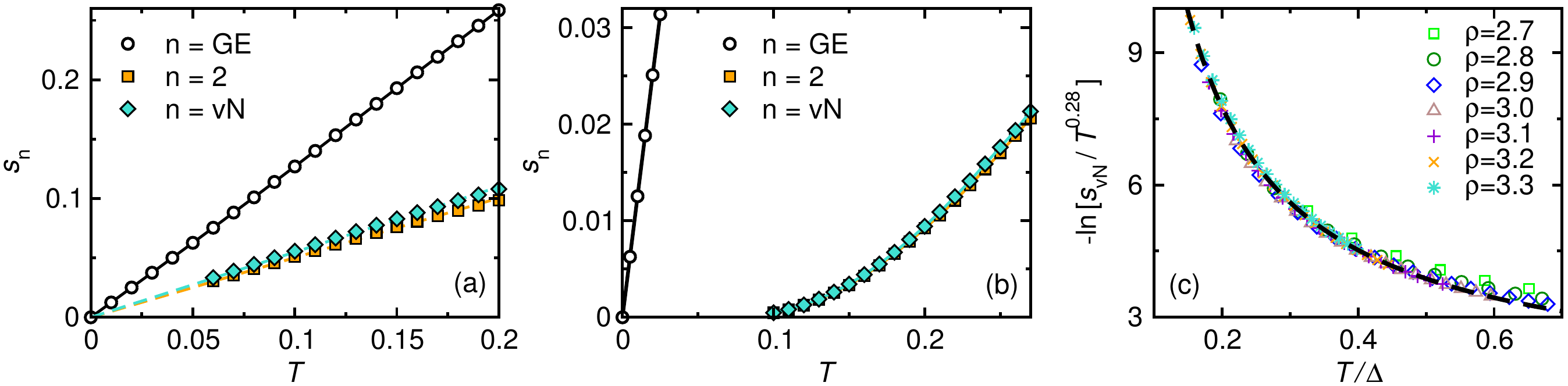}
\caption{Average eigenstate entanglement entropy density $s_{\rm n}$, as obtained from fits to Eq.~\eqref{def_S_T_fit}, and the thermodynamic entropy density $s_{\rm GE}=\lim_{R\to\infty} S_{\rm GE}/(2R)$. The entropy densities are plotted as functions of temperature for: (a) $\rho=2$ and (b) $\rho=3$. (c) Data collapse of the von Neumann entanglement entropy density for different values of $\rho \gtrsim \rho_c$. The symbols depict the numerical results. The dashed lines in panel (a) are linear fits according to Eq.~(\ref{def_S_T_linear}), while the solid lines in panels (a) and (b) are guides to the eye. The dashed line in panel (c) is the function $1.295 (\Delta/T) + 1.288$.} 
\label{fig6}
\end{center}
\end{figure*}

Figures~\ref{fig6}(a) and~\ref{fig6}(b) show the average eigenstate entanglement entropy densities $s_{\rm n}$, together with the corresponding thermodynamic entropy density $s_{\rm GE}=\lim_{R\to\infty} S_{\rm GE}/(2R)$, as functions of temperature at low temperatures. The thermodynamic entropy density exhibits a linear increase with $T$ for all values of $\rho$, analogous to the one observed for free fermions in a homogeneous lattice. This is a consequence of the system being gapless independently of the value of $\rho$.

On the other hand, the average eigenstate entanglement entropy densities exhibit a qualitatively different behavior depending on whether $\rho$ is smaller or larger than $\rho_c$.
For $\rho<\rho_c$, as $s_{\rm GE}$, they exhibit a linear increase with temperature
\begin{equation} \label{def_S_T_linear}
s_{\rm n}(T,\rho) = \alpha_n(\rho) \,T\,, \hspace{0.3cm} s_{\rm GE}(T,\rho) = \alpha_{\rm GE}(\rho)\,T\,,
\end{equation}
with $\alpha_n(\rho) < \alpha_{\rm GE}(\rho)$. For $\rho > \rho_c$, however, the average eigenstate entanglement entropy densities exhibit an activated-like behavior that is absent in $s_{\rm GE}$. We fit the results for the von Neumann entanglement entropy density with the ansatz
\begin{equation} \label{def_s_activated}
s_{\rm vN}(T,\rho) = a\, T^\zeta \exp\left[-b\, \left( \frac{\Delta(\rho)}{T}\right)^\gamma \right] \,,
\end{equation}
where $\Delta(\rho) = \rho - \rho_c$. The analysis described in Appendix~\ref{app4} reveals that the optimal exponent of the temperature dependent prefactor is $\zeta = 0.28$. We then evaluate the rescaled function $\tilde{s}_{\rm vN} = -\ln[s_{\rm vN}/T^\zeta]$, see the symbols in Fig.~\ref{fig6}(c), which results in excellent data collapse for different values of $\rho$ when $T/\Delta \lesssim 0.5$. The dashed line in Fig.~\ref{fig6}(c) is a simple algebraic function $b\, (\Delta/T) - \ln a$ [i.e., setting $\gamma=1$ in Eq.~(\ref{def_s_activated})], which is an excellent match to the numerical results.

It is expected that, in general, $s_{\rm n}$ is smaller than $s_{\rm GE}$ at all temperatures and characteristic densities. This follows a recent study~\cite{vidmar_hackl_17}, in which it was proved that $s_{\rm vN} < s_{\rm GE}$ for identical bipartitions of translationally invariant fermionic quadratic models at infinite temperature. Our results for $\rho > \rho_c$ in trapped systems show that the average eigenstate entanglement entropy density $s_{\rm n}$ and the thermodynamic entropy density $s_{\rm GE}$ can actually exhibit qualitatively different behavior (which was not the case in Ref.~\cite{vidmar_hackl_17}). The former exhibits activated-like behavior while the latter is linear in temperature. Furthermore, the data collapse in Fig.~\ref{fig6}(c) shows that $\Delta$ acts as a sort of gap for the eigenstate entanglement entropies in trapped systems (it determines the width of the band-insulating domain in the center of the trap). This despite the fact that that there is no energy gap in the single-particle spectrum of the Hamiltonian~(\ref{def_H}), as shown in Fig.~\ref{fig1}(a), and, consequently, in the many-body energy spectrum. Hence, at low temperatures, the average eigenstate entanglement entropies exhibit fingerprints of the formation of the band-insulating domain, i.e., of the local quantum phase transition, and could be used to locate it.

\section{Summary} \label{sec5}

We used quantum information measures to study spinless fermions in one-dimensional lattices in the presence of a harmonic trap. In contrast to most studies of entanglement entropies, which focus on small subsystems of ground states, here we considered a bipartition of the system in two equal parts and studied ground states, excited states, and finite-temperature mixed states. 

In the ground state, we showed that the entanglement entropies scale as $\ln N$ at small characteristic densities~\cite{vicari12}, while they vanish after the emergence of the band-insulating domain (as expected from the fact that one has two disconnected metallic domains). More importantly, we showed that a scaling analysis of the entanglement entropies in finite systems allows one to accurately determine the critical characteristic density for the formation of the band-insulating domain. Hence, ground-state bipartite entanglement entropies are excellent order parameters to describe the characteristic-density driven local quantum phase transition in those systems.

For low-temperature thermal states, we showed that (i) a sharp peak in the thermodynamic entropy, associated with a divergence of the single-particle density of states at the Fermi energy corresponding to $\rho_c$, signals the critical point, and (ii) the mutual information exhibits a behavior that is qualitatively similar to that of the entanglement entropies in the ground state, namely, it is nonzero below the local quantum phase transition and very small above it. Increasing the temperature destroys this contrast. 

Finally, we systematically studied excited eigenstates. We showed that their average entanglement entropy scales linearly with $N$, i.e., the overwhelming majority of the excited states exhibit a volume law, but the prefactor is always smaller than that of the thermodynamic entropy. This complements many recent works that have studied entanglement entropies for lattice bipartitions in which the smaller subsystem is not a vanishing fraction of the system ~\cite{santos_polkovnikov_12, deutsch13, storms14, beugeling15, garrison15, nandy_sen_16, dymarsky_laskhari_17, fujita_nakagawa_17, vidmar_rigol_17, huang_17, lu_grover_17, chen_ludwig_17}, and, in particular, the works dealing with quadratic models~\cite{vidmar_hackl_17, storms14, nandy_sen_16}. Our main finding for excited states is that, as functions of the temperature, the average eigenstate entanglement entropy densities behave very differently for $\rho<\rho_c$ (linear increase with the temperature) and $\rho>\rho_c$ (activated-like behavior). Hence, at low temperatures, they carry the fingerprints of the local quantum phase transition.

\section{Acknowledgments}

We acknowledge discussions with R. Modak and V. Alba. This work was supported by the NSF, Grant No.~PHY-1707482.

\appendix

\section{Site occupations for $\rho \leq \rho_c$} \label{app0}

\begin{figure}[!h]
\begin{center}
\includegraphics[width=0.99\columnwidth]{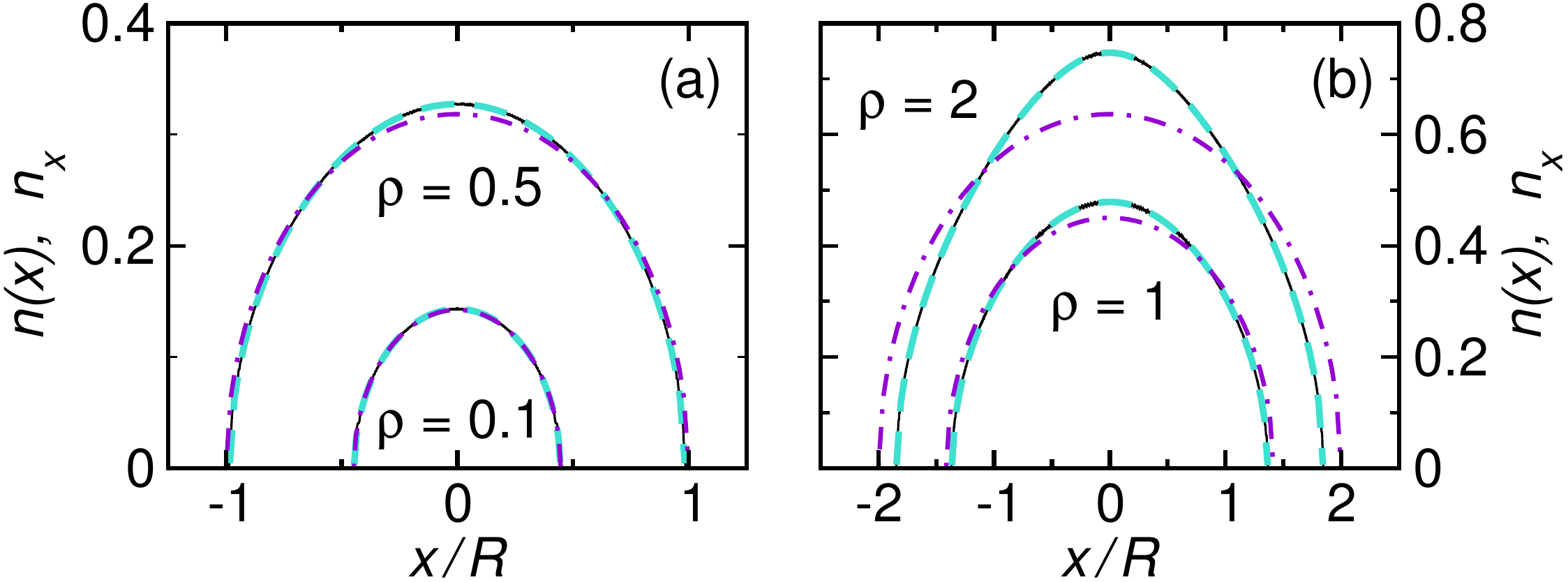}
\caption{Site occupations as functions of $x/R$ for different values of $\rho$. Solid lines are exact numerical results $n_x = \langle \hat n_x \rangle$ (with integer $x$ and $R=500\pi$), and the overlapping dashed lines are the LDA results $n(x)$ from Eq.~(\ref{def_n_x_0}). The dashed-dotted lines are the results for the continuum (limit $\rho \to 0$ in a lattice), given by Eq.~(\ref{def_n_sc}).} 
\label{figapp0}
\end{center}
\end{figure}

For $\rho \leq \rho_c$, the site occupations $n(x)$ within the LDA are given by Eq.~(\ref{def_n_x_0}). We obtain the relation between $\rho$ and $\mu_0$ from the equation $\rho = \int_{-x_0}^{x_0} n(x)dx \, /R$. It yields
\begin{equation} \label{def_rho_mu0}
 \rho = \frac{4 \sqrt{2-\mu_0}}{\pi} \left( E\left[ \frac{\mu_0 +2}{ \mu_0 - 2} \right] - K\left[ \frac{\mu_0+2}{ \mu_0 - 2} \right] \right) \, ,
\end{equation}
where $K$ and $E$ are the complete elliptic integrals of the first and second kind, respectively. For any given $\rho$, we calculate $\mu_0$ numerically from Eq.~(\ref{def_rho_mu0}). $\mu_0$ is needed to determine $n(x)$. The agreement between $n(x)$ (dashed lines) and the numerically exact site occupations $n_x$ (solid lines), shown in Fig.~\ref{fig1}(b) and Fig.~\ref{figapp0} as a function of $x/R$, is excellent. We also compare both results to those in the continuum limit (see, e.g., Refs.~\cite{xu_rigol_15, dubail_stephan_17}), which can be obtained from Eq.~(\ref{def_rho_mu0}) by expanding $\mu_0$ around $-2$. This yields a simple relation $\rho = 1 + \mu_0/2$ and the semicircle particle distribution
\begin{equation} \label{def_n_sc}
 n_{\rm sc}(x) = \frac{1}{\pi} \sqrt{2\rho - \left( \frac{x}{R} \right)^2} \, .
\end{equation}
The results for $n_{\rm sc}(x)$ are shown as dashed-dotted lines in Fig.~\ref{figapp0}. They are nearly indistinguishable from the exact ones up to $\rho \lesssim 0.5$.

\section{Numerical evaluation of Eq.~(\ref{def_S_T})} \label{app1}

We describe how the average eigenstate entanglement entropy $\bar S_{\rm n}$, defined in Eq.~(\ref{def_S_T}), is evaluated numerically. We focus on systems with a fixed particle number $N$ (canonical ensemble averages), while a comparison with grand canonical ensemble averages is presented in Appendix~\ref{app2}.

In a system of $N$ spinless fermions in a one-dimensional lattice with $L$ sites, the total number of many-body energy eigenstates is $\binom{L}{N}$. Among all the energy eigenstates, we systematically find those ($|m\rangle$) with a relative weight $\exp{\left[-(E_m-E_{\rm GS})/T\right]} > \exp{[-\Lambda]}$, where $E_{\rm GS}$ is the ground-state energy, $E_{\rm GS}=\sum_{i=1}^N{\varepsilon_i}$, and $\Lambda$ sets the numerical accuracy. Note that the single-particle eigenenergies ${\varepsilon_i}$ are ordered such that ${\varepsilon_i}<{\varepsilon_{i+1}}$.

\begin{figure}[!t]
\begin{center}
\includegraphics[width=0.99\columnwidth]{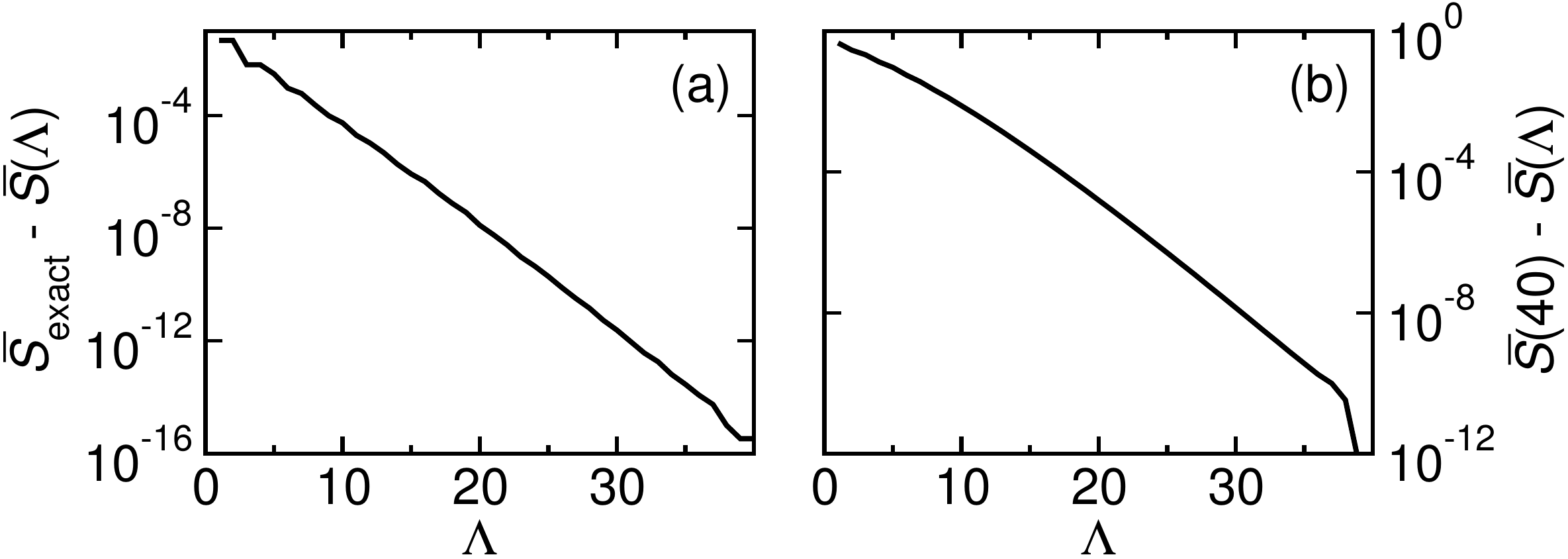}
\caption{Accuracy of the truncation used to evaluate the average von Neumann eigenstate entanglement entropy in Eq.~(\ref{def_S_T}), for $\rho=2$ and $T=0.1$. (a) Difference between the exact average $\bar S_{\rm exact}$ and the result after the truncation $\bar S(\Lambda)$, for $R=5$ and $L=24$. (b) $\bar S(40) - \bar S(\Lambda)$ for $R=20$ and $L=150$.} 
\label{figapp1}
\end{center}
\end{figure}

The accuracy of such a truncation scheme is studied in Fig.~\ref{figapp1}(a) for the von Neumann entanglement entropy $\bar S_{\rm vN}$, for $\rho = 2$, $T=0.1$, and $L=24$. The results show that the difference between the exact average $\bar S_{\rm exact}$ and the average after truncating the sum $\bar S(\Lambda)$ decreases exponentially with $\Lambda$. For larger systems ($L \gtrsim 100$), however, it is numerically impossible to obtain the exact averages. In Fig.~\ref{figapp1}(b), we report the difference between $\bar S(\Lambda)$ and $\bar S(\Lambda=40)$, for $\rho = 2$, $T=0.1$, and $L=150$ ($\Lambda=40$ is about the largest cut off we can consider for that system size). The differences can again be seen to decay nearly exponentially, and to be very small ($\sim10^{-8}$) for $\Lambda = 30$. This is the value of $\Lambda$ used to obtain the results reported in Figs.~\ref{fig5},~\ref{fig6},~\ref{figapp2}, and~\ref{figapp3}.

\section{Canonical vs grand canonical averages} \label{app2}

\begin{figure}[!t]
\begin{center}
\includegraphics[width=0.99\columnwidth]{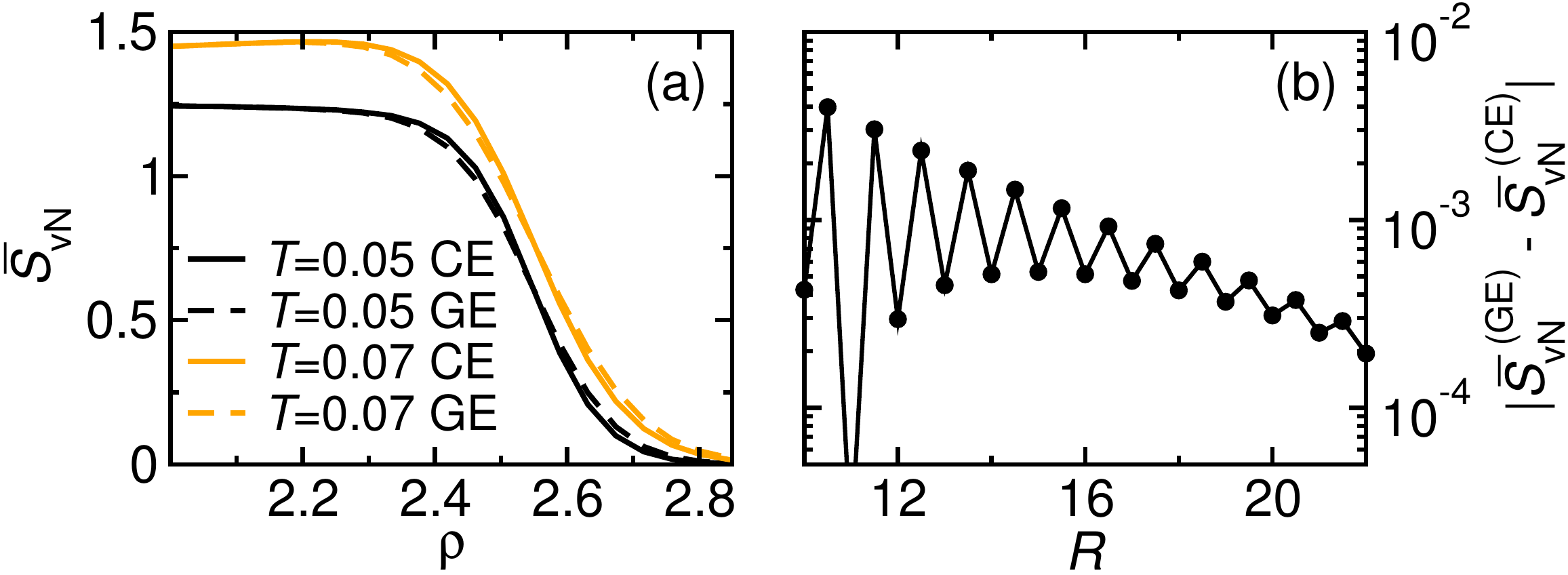}
\caption{Comparison between the canonical and grand canonical averages of the von Neumann eigenstate entanglement entropy $\bar S_{\rm vN}$. The former is calculated by using Eq.~(\ref{def_S_T}), while the latter one is calculated by using Eq.~(\ref{def_S_T_GE}). (a) Results for $R=7.5\pi$ as a function of $\rho$, for two temperatures. (b) Absolute difference between $\bar S_{\rm vN}^{\rm(CE)}$ and $\bar S_{\rm vN}^{\rm(GE)}$, at $\rho = 2$ and $T=0.1$, as a function of $R$.} 
\label{figapp2}
\end{center}
\end{figure}

We extend the average eigenstate entanglement entropy in Eq.~(\ref{def_S_T}), which corresponds to a canonical calculation, to a grand canonical average
\begin{equation} \label{def_S_T_GE}
\bar S_{\rm n}(T) = \frac{\sum_N\sum_{m(N)}{S_{\rm n}(m)e^{-(E_m-\mu N)/T}}}{\sum_N\sum_{m(N)}{e^{-(E_m-\mu N)/T}}}\, ,
\end{equation}
where the chemical potential $\mu$ is determined so that the average particle number is $N$. In contrast to the canonical average, one needs to perform an additional sum over sectors with different particle numbers $N$ and many-body eigenstates $\{|m(N)\rangle\}$. This increases significantly the computation time.

Figure~\ref{figapp2}(a) compares the canonical and grand canonical averages of the von Neumann entanglement entropy $\bar S_{\rm vN}$, plotted as functions of $\rho$. The results are very close for all $\rho$. Figure~\ref{figapp2}(b) shows the difference between the two averages upon increasing the system size (or, equivalently, $R$) for $\rho = 2$. As expected, the differences decrease with increasing system size, and are expected to vanish in the thermodynamic limit ($R\to\infty$ while keeping $\rho$ constant). Because of this, we only report results for canonical averages in Figs.~\ref{fig5},~\ref{fig6}, and~\ref{figapp3}.

\section{Volume-law scaling of the average eigenstate entanglement entropy} \label{app3}

\begin{figure}[!h]
\begin{center}
\includegraphics[width=0.99\columnwidth]{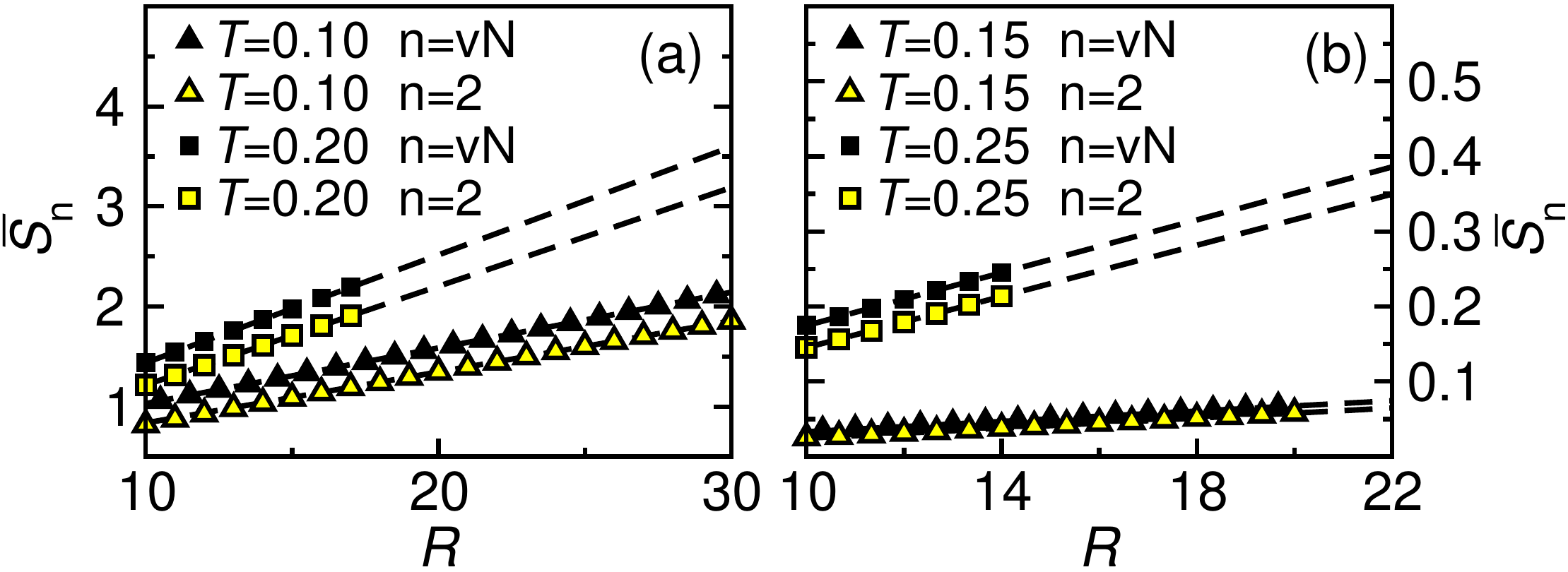}
\caption{Volume-law scaling of the average eigenstate entanglement entropies. (a) $\rho = 2$ and (b) $\rho = 3$. Symbols display the von Neumann entropy $\bar S_{\rm vN}$ and the second Renyi entropy $\bar S_{2}$ for different temperatures. Dashed lines are linear fits using the ansatz in Eq.~(\ref{def_S_T_fit}).} 
\label{figapp3}
\end{center}
\end{figure}

Figure~\ref{figapp3} shows that $\bar S_{\rm n}$ increases linearly with $R$ at fixed temperature, as stated in Sec.~\ref{sec4b}. Note that results are reported for $\rho=2$ ($\rho<\rho_c$) and $\rho=3$ ($\rho>\rho_c$) for different temperatures. This justifies the use of the ansatz in Eq.~(\ref{def_S_T_fit}) to extract the average eigenstate entanglement entropy density $s_{\rm n}$ reported in Fig.~\ref{fig6}.

\section{Average eigenstate entanglement entropy for $\rho > \rho_c$} \label{app4}

Figure~\ref{fig6}(b) shows an activated-like behavior of $s_{\rm n}(T,\rho)$ as a function of $T$, for $\rho > \rho_c$. We fit the numerical results to the ansatz in Eq.~(\ref{def_s_activated}) with $\zeta=0.28$. To determine this optimal value of $\zeta$, we did as follows: We calculated $\tilde s_{\rm vN} = -\ln [s_{\rm vN}/T^{\zeta}]$ vs $\Delta/T$ for different values of $\zeta$ in the interval $\zeta \in [-0.5,1]$. For a fixed value of $\zeta$, we fit the results for $\tilde s_{\rm vN}$ (obtained for seven values of $\rho= \{2.7, 2.8, \dots, 3.3 \}$) with a high-order polynomial $\sum_{n=0}^6 a_n (\Delta/T)^n$ in the regime $\Delta/T \leq 0.5$. We then calculated the sum of squares of the differences between $\tilde s_{\rm vN}$ and the fitted polynomial. This sum was found to have a minimum at $\zeta = 0.28$.

\bibliographystyle{biblev1}
\bibliography{references}

\end{document}